# CHANGE MANAGEMENT: IMPLEMENTATION AND BENEFITS OF THE CHANGE CONTROL IN THE INFORMATION TECHNOLOGY ENVIRONMENT

Paulo Roberto Martins de Andrade[1], Adriano B. Albuquerque[2], WeskleiDourado Teófilo[3], Fátima Aguiar da Silva[3]

[1]Graduate Studies in Computer Science, University of Regina, Regina, SK, Canada
[2] Universidade de Fortaleza, Fortaleza, CE, Brazil
[3]FaculdadeEstácio FIC, Fortaleza, CE, Brazil

## ABSTRACT

*In the competitive environment, companies have given increasing importance to the IT sector and the resources it delivers as strategic. As a result, IT becomes a living being within the company. This sector is being subject to continuous changes in this scenario. These changes can occur within the own IT sector or whether IT to other sectors of the company. For both scenarios, it is important to have a good change control to avoid unnecessary trouble and expense. This paper aims to show through a case study, the benefits and results obtained with the implementation of a process of managing and controlling changes in the information technology environment of a large government company in Brazil.*

## 1. INTRODUCTION

We live in a time when technology advances faster and faster, and with these advances and changes in technology, corporations have to keep pace to be competitive in the increasingly computerized and demanding market. Exactly at this point enter the changes made in systems and network infrastructure.

In the past, the common knowledge said that the IT department of the organization need to be aligned to the business, but today the IT is already part of the business. Both, the IT and the Business' Department, are two areas that should be "integrated". With that in mind, it is necessary that the IT department is prepared to support and ensure timely, delivery of IT services required for the company's business and can also add value to the corporation [17].

With some of its foundational texts now decades old, IT change management is one of the most studied and utilized elements of the systems and infrastructure management discipline. With so much academic thinking and so many years of practical application behind it, one would think that the effective management of change would now be commonplace in IT organizations and the companies they serve. Surprisingly, the opposite it true. Author and IT Service Management expert, Harris Kern, reports that in a recent survey of 40 corporate IT infrastructure managers a surprising 60% admitted that their processes to handle change are not effective in communicating and coordinating changes occurring within their production environment. Among the key findings of the study: Not all changes are logged 95%; Changes not thoroughly tested 90%; Lack of process enforcement; Poor change communication and dissemination; Lack of centralized process ownership 60%; Lack of change approval policy 50%; Frequent change notification after the fact 40% [9].

More than an overall indictment of poor process implementation, these findings illustrate the extent to which organizations rely on a change control process to handle IT-related changes.

Changes can be welcomed and usually they happen to have benefits. What is needed is to manage them the right way to not to be surprised negatively on the results. For every change has an expectation that must be the closest possible to the final result of the change.

In order to have a management and control appropriate changes, the main aim of this paper is to define a process to carry out the treatment of changes in the information technology environment, deploying and generating benefits, because it is an extremely important environment for companies today. We must select tools to be used, define the stakeholders, ways of communication and approval roles. Thus we can define a process of change, starting from the request, its type, its approval, its execution and the lessons learned. With this formalized process, we can have the benefits expected as a higher rate of success in the changes, greater traceability, risk mitigation, among several other potential benefits with the use of the process.

This paper delivery information about the sources of change and one some concepts of changes in the Project Management. Next, the paper shows us the previous scenario overview. After, we have a case study (with the problems founded, the benefits expected and the difficult in the implantation. In the next we have the results of the case study. At last, the paper gives us the conclusion.

## 2. MORE ABOUT CHANGES

### 2.1. Sources of Changes

In our lives or in our projects we should always be prepared to deal with and make changes because they always occur. In the development of software applications, we recommend the use of good practices related to Software Engineering. For IT Governance there are several frameworks of best practice, one example would be the COBIT with BAI06 Manage Change process [10].

The COBIT is an IT governance framework maintained by ISACA (ISACA). It is a reference model and is not a ready solution. In other words, it needs to adapt it to the business to meet the needs, and align the information technology sector to the business and can be used only a few controls that seek to improve the processes that comprise the IT governance of each company [10].

There are several causes that give rise to the changes, we can highlight the compliance with laws and regulations such as PCI-DSS (Payment Card Industry - Data Security Standard) [11], SOX (Sarbanes Oxley Act) [16], needs preventive measures in the IT environment, external events of which the IT has to comply in determined period, need for improvements in systems and infrastructure, senior management requests (strategic changes), but the highest incidence of changes in the IT environment noted in the case study in this article are those that originate from problems. In this case, the change is the answer to the problem. Another important part of the change comes from direct demands from any interested party.

## 2.2. Concepts in the view of Project Management

The Project Management Institute (PMI) is a non-profit organization that combines project management professionals around the world integrating more than 2.9 million members in nearly every country through research and education. Its central office is located in Pennsylvania (United States) [4].

The PMBOK identifies a subset of knowledge in project management, which is generally recognized as good practice, and as a result, used by the Project Management Institute (PMI), which defines some concepts down on change management [12]:

"A change request is a formal proposal to modify any document delivery, or baseline". (PMBOK, 2013, p. 85);
"Perform Integrated Change Control is the process of reviewing all change requests, approving changes and managing changes being made deliveries, organizational process assets, project documents and project management plan, and report the disposition of thereof". (PMBOK, 2013, p. 94).

The PMBOK describes a process for the integrated change control with a focus on project management, in which he mentions the main inputs, tools and techniques, and outputs of the process as shown in Figure 1.

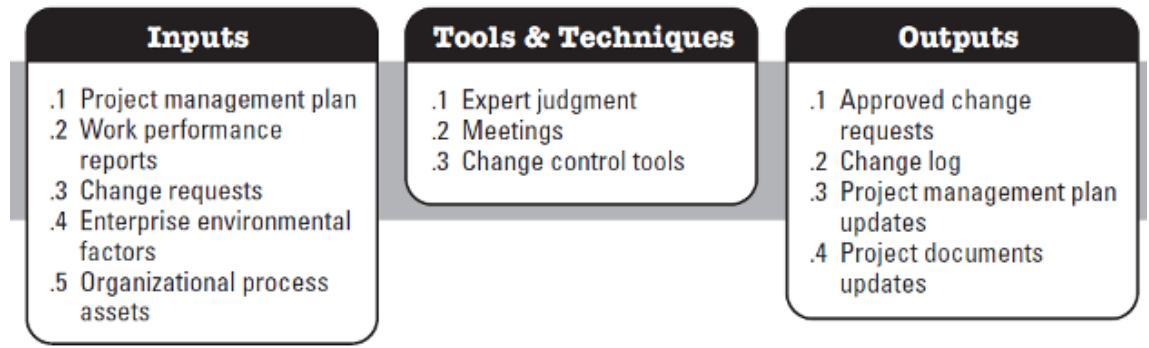

Figure 1. Perform Integrated Change Control Process
(source: PMBOK 5ed. P. 94)

This process can be adaptable to each company, environment or need, since the focus of the PMBOK guide is for project management environment.

There is also another project management method called PRINCE2 is recognized worldwide through the dissemination of good practice in project management, in which speech and deals with the importance of using change management [1].

PRINCE2 is a non owner who is emerging worldwide as one of the most accepted methods for project management and is much used mainly in European countries. It is truly flexible and can be applied to any project or company, regardless of scale, type, organization, geography or culture. Addresses project management with four integrated elements: principles, themes, processes and project environment. One of its themes is exactly "change" and has the purpose to identify, assess and monitor any changes [2].

The fact that the PRINCE2 have a specific area for change management help in the success of the project, and the changes made. It is evident that the changes will always occur and that

25

communication of them is very important for effective implementation, because in this way reduces the probability of failure, time and cost, and thus fall into important and fundamental point that is customer satisfaction and success the purpose of the change [1].

The PMI and PRINCE2 (two of the biggest disseminators organs of good practice in project management, world-renowned) address change management emphatically as an important point for the success of projects, stating on multiple pages of your guides, good practices that can be used [13].

The reason to talk specifically about that topic changes in project management is to be the area where most uses the concepts, and where it is more widespread and used the change management process. This process can be easily adapted for information technology, using these best practices in order to achieve better results in the implementation and use of change management [3].

## 2.3 The importance of the change management

Organizations change to win. And this gain is not always financial. Sometimes the aim is to just improve interpersonal relationship, hasten the progress of internal processes or stimulate the involvement of employees in a new project. The possibilities are many and invests up time and money to achieve the goals. However, in many organizational changes the desired result is far away. And that does not happen by accident.

When a company is created, its founders have beliefs, values and assumptions related to the way of how your business should work. If the company prospers, it is natural to hire people who share the same 'feelings'. These beliefs, values and assumptions are represented by the norms of behaviour and attitudes accepted by the organization. They are taught as "the correct way it should be done things around here."

It can not be said, in an isolated manner, if a culture is good or bad. However, it is known that a culture is good if it is consistent with the company's strategy. For example, companies that depend on the creativity - such as those of advertisement, for example - may not have hierarchies and rigid processes; moreover, it does no longer true in chemical and pharmaceutical companies. Anyway, for us it is extremely important to be always attentive to the evolution of companies, so we realize that the same culture that led to success may no longer be useful to prosper in the future. And it's hard for some managers see themselves the need for change.

Many managers fear that their subordinates resist the changes. Therefore, they do not play the role expected of them in any transformation process. Culture is created by the leaders and therefore, if we need to change, this change must begin with the leaders. It is they who must create a sense of need and urgency, to establish the vision and communicate it clearly and honestly.

A change may require changes in existing structures, new skills needs, work load shifting, etc. All of these impacts should be managed effectively so that the transformation can occur. The transition between the current state and the future state should be monitored. Support new processes, contingency planning and treatment of risks, plans and support change issues must not be neglected. Training is an extremely important aspect of the process, but often is mishandled during a transformation. For this reason, it is responsible for the failure of much of the organizational changes.

As can be seen, the success of organizational change processes is not so simple, requires attention and support professionals. The fact is that companies with proper professional support are evolving and achieved the desired results and this finding has a valuable significance for businesses, especially for those who are facing the challenge of taking its first steps into the future.

## 3. PREVIOUS SCENARIO OVERVIEW

### 3.1 Problems and benefits expected

The scenario in which we performed the case study of this article it is a large government company, where we diagnose various problems related to changes in the information technology environment, such as systems, servers and network infrastructure. Routine problems were:

- Unplanned outages of services. The unavailability of services mostly occurred during business hours, impacting users and customers of the company, thus affecting the credibility of the information technology sector.
- Difficulty in tracing changes in the environment for troubleshooting. As the changes were made without any control we did not know what had changed in the environment, even a short time ago, making diagnosis difficult to solve problems, often resulting in increased downtime.
- Risks unmitigated in carrying out the changes. As the risks were not analyzed and treated in the best way, the impact of changes often were not as expected.
- Rework. A large percentage of the changes generated rework.
- Low success rate in the changes.
- Financial losses not measured, especially on the availability of systems and services.

These are some of the many factors that contributed in the interest of implementing a change control process in the company. According to the market more and more competitive, and especially following the controls Framework COBIT and PMBOK, these good practices also would make other protruding factors to consider the real need to implement a process of integrated control of changes in management of enterprise information technology, seeking the benefits when using this process [6].

The main expected benefits that motivated the process of implementation were:

- The better alignment of IT services with business. The changes will be filtered and prioritized as needed for business.
- Increased visibility into the changes by exercising greater control over the execution of change, reducing the negative impact of change.
- The risk analysis to prevent the service becomes unavailable due to failures.
- Better assessment of the cost-benefit ratio of the change before it is implemented.
- Reduction of cost, time and rework.
- More stable services and increase of user productivity.

These are some of the expected benefits of the process of implementation. The problems and benefits were analysed along with IT teams, managers and key stakeholders, thus coming to the conclusion that it would be quite feasible the implementation of an integrated control changes in the information technology sector.

### 3.2 Difficult of the Change Management Process implementation

During the implementation process were expected difficulties due to the sector's culture and its employees, being a paradigm shift because there were resistance from some people, especially for thinking that with this process, increase bureaucracy and this way, possible problems in the information technology sector, where more and more is expected quick answers to problems and incidents could take even longer to resolve.

Andrade defines organizational culture as a set of beliefs, traditions, values, written and unwritten rules that can boost, speed, weaken, delay, facilitate, commit, hinder or prevent changes and performance of organizations [3].

According Chiavenato, the change is the main characteristic of contemporary times, being an essential aspect of creativity and innovation in organizations today [7].

Barbara mentions that: "Changes happen when people are prepared for the same, however, they find it difficult to break paradigms, changing behaviour and changing their attitudes, because changes imply something new and they are still attached to the previous archetype, causing uncertainty and being affected psychologically, causing fear and consequently generating resistance." [5].

The main challenge of the deployment and use of change control process is to provide an organizational environment propitious to innovation that adds value to the company and to the information technology sector. In them will be developed and implemented projects and activities that generate changes, working mostly in dimensions' people and results.

The process creation, dissemination, definition of approval levels and roles were performed as expected. The greatest difficulty really was a culture change so that it could follow the process, especially at the beginning when sometimes tried to put the blame problems unrelated to the process in the process itself, being something new [15].

## 4. CASE STUDY

First of all, we need a formal process control changes that would fit as close as possible to the culture and reality of the IT sector and the company. We defined with all stakeholders in the process that any changes in the information technology environment should be requested, and formally managed through the integrated control of process changes.

Through the definition that any changes would use the process, we saw the need to categorize the changes because there were cases where differential treatment would be needed for certain changes, aiming to make the process as effective as possible. To this categorization were defined three types of changes:

Standard Change (Routine): These are pre-approved changes. This means that new requests are automatically approved and occur in periods intervals previously defined.

Planned changes: Are the changes that run the normal flow within the process.

Emergency changes: Are the changes that may occur outside the standard interval, only depending on the release of the immediate supervisor for immediate execution, are generally reactive changes to correct problems that are impacting or may impact the business.

International Journal of Advanced Information Technology (IJAIT) Vol. 6, No. 1, February 2016

Any changes in the information technology environment must go through the process, being necessary to open a change request, which was made available on the company's Intranet a form to fill the requests. The Figure 2 show this form and in the Figure 3 is possible to see the tracking of each request in the system. Both figures are in pt-BR (Portuguese – Brazil).

Figure 2. Form of change request

Figure 3. Tracking screen for the Change Requests

On the form, you are asked to enter the following information:

- What is the purpose of the change?
- Start date and completion of the change.
- What will assets be affected by the change?
- What will items be impacted by the assets affected by the change?
- What is the complexity of change?
- Further details on the change.
- Which the implementation experience of change?
- Tests performed to carry out the change.
- What is the return plan for change?

The roles were defined for analysis and approval of the changes as described below:

- Requester: Any employee of IT sector that want to make the change, and the requester also mostly responsible for the execution and implementation of the same, then the execution of the change should be completed by the requester or designed responsible.
- Governance and IT Risk Team: Responsible for evaluate the request, categorize the change, send the flow of approvals, conduct communication to stakeholders, update the documentation and record the closure of the change.
- Change Team: Responsible for assessing the feasibility of change, checking the necessary approval levels for the request.
- Immediate supervisor: They are the persons responsible for the approval of the changes, usually managers and members of business teams impacted the change.

After requesting submit the form changes, the application will be reviewed. If necessary, will be asked to charge to carry out possible corrections in your request. Then the change will be categorized and will cover the flow according to its categorization also being examined its feasibility, and responsible for approval. After the change is formally approved for execution, the team communicates to all parties affected by the change. This communication become more transparent to all the actions taken by TI, increasing their credibility with employees. Shortly after the execution of change, should be completed a post-implementation form with the questions shown below:

- Was the purpose of the change reached?
- Did the change occur within the expected interval?
- Was there any incident arising from the change?
- Was it necessary to trigger the return plan (rollback)?
- Will be necessary upgrade documentation?
- Other comments.

After the post-implementation form is submitted, if necessary, documentation will be update, and thus the closure of the change will be recorded.

## 5. RESULTS

To be able to measure the results of the integrated control process changes, create a performance indicator (KPI) to monitor and improve the process, using practices of continuous improvement. The indicator was set to be four times a year, consisting only the final result of the change, which may be success or failure. The changes considered successful, briefly, are the ones that occurred within the expected time period and had reached their ultimate goal, otherwise the change is classified as failed.

After the first six months of 2015, we have recorded 200 requests for changes through the form found on the intranet. They all went through in the process designed and previously mentioned. Here entered changes relating to update systems, update hardware, implementation of new systems (to make sure that deployment could affect some other IT assets), emergency changes, among other types. With AJUSA the process, only 17 changes obtained a failure during its implementation, it is necessary to enable the return plan. Figure 4 illustrates this data.

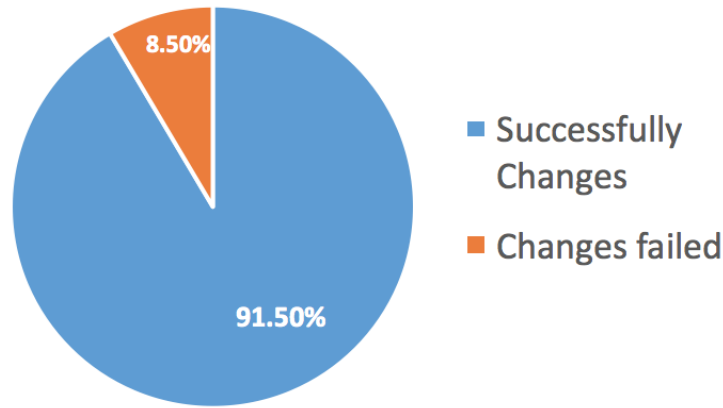

Figure 2. Collected result from the first six months of 2015

With the formalization and implementation, the process has matured gradually until you can be part of the IT team culture. It was observed the increase of environmental availability through monitoring reports after the process of implementation. The changes have achieved their goals almost entirely. The risks are mitigated to be obtained minimal impact to the environment in the changes.

A major benefit noted by all the company's employees was communication. Earlier the systems and servers were unavailable without prior knowledge of users, making it stressful communication with the IT sector. With the process implementation, the system's stops are scheduled and even emergencies, are communicated to all interested parties, which made up the credibility of the IT sector in the company. With the creation of the KPI to monitor the change control process there was come in quite satisfactory results, as we saw in the figure 2.

## 6. CONCLUSIONS

As mentioned throughout this paper, the use of change management is becoming increasingly essential for the Information Technology sector, and for businesses, helping and supporting the achievement of strategic planning, targets, and indicators. With the highly competitive market that we live in today, and growth and alignment of information technology with business, adding more value to the enterprise processes such as change management, can be a differentiator so that you can excel in current market scenario. The implementation of the change management process is not complicated, but you should not expect its implementation from day to night, we need to adapt the process to the culture and maturity of the company, requiring dedication, commitment and effort everyone involved, so that we can break resistance and paradigms related to this change with the use of the process.

It is observed although the implementation of a change management process brings many benefits as cited throughout this article. It is a process that does not require much financial investment, making it highly feasible and recommended for use. For the process, the main barrier to successful were the people. With time and the use of the process is perceived the change by the involved culture, making it easier deployment and standardization of other processes in order to use of best market practices in order to add the most value to the business and making a difference in today's competitive market.

## ACKNOWLEDGEMENTS

The CNPq (Conselho Nacional de Desenvolvimento Científico e Tecnológico - "National Counsel of Technological and Scientific Development") and the Ceará water and sewage company support this work.

## AUTHORS

**Paulo Roberto Martins deAndrade** is a PhD Student in the University of Regina andAssociated Professor in the "Estácio FIC". Actually, his focus is in Project Management. He has 11 years of experience in IT, with some certifications, like PMP and Cobit. He has a Master in Applied Informatics, MBA in Project Management and MBA in IT Management.

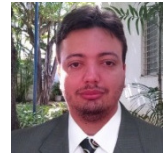

**Adriano Bessa Albuquerque** has Doctorate in Systems Engineering and Computing and a Master in Applied. He has experience in computer science with emphases in software engineering. Actually he is a professor in the University of Fortaleza and he also Acts as implementer and official evaluator of software maturity models: CMMI and MPS.BR.

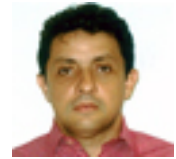

**WeskleiDouradoTeófilo**has a degree in Computer Networks, MBA in IT Governance and MBA in Project Management. He has more than 15 certifications, among which are, PMP, COBIT and CompTIA Security +. He has been working for over 8 years in the field of Information Technology, with an emphasis on IT Governance, Project Management and Information Security.

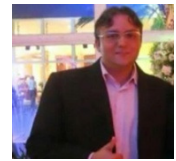

**Fátima Aguiar da Silva**has a degree in Business Administration and MBA in Project Management. She has certifications likeCAPM, Cobit and ITIL.Shehas been working for over 6 years in the field of Process Management, with an emphasis on IT Governance and Project Management.

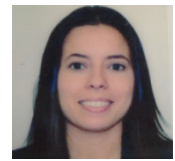